\documentclass[letterpaper]{article}

\usepackage{setspace}
\usepackage[T1]{fontenc}
\usepackage[utf8]{inputenc}
\usepackage{siunitx}

\usepackage{xcolor}

\usepackage{soul}

\usepackage{geometry}
\usepackage{setspace}

\usepackage[style = chem-acs]{biblatex}
\usepackage{graphicx}
\usepackage{float}
\newfloat{scheme}{htbp}{los}
\floatname{scheme}{Scheme}
\floatname{chart}{Chart}
\newfloat{graph}{htbp}{loh}

\usepackage{chemformula} 
\usepackage[version = 4]{mhchem} 

\usepackage{authblk}
\author[1,2]{Isaac A. Lagaud}
\author[1]{Justin Boddison-Chouinard}
\author[1,2]{Antoine Labbé}
\author[1]{Ayse M. Aygar}
\author[1]{Alex Bogan}
\author[1]{Li-Lin Tay}
\author[3]{Takashi Taniguchi}
\author[4]{Kenji Watanabe}
\author[5]{Aljoscha Söll}
\author[5]{Zden\v{e}k Sofer}
\author[2,6]{Adina Luican-Mayer}
\author[1]{Louis Gaudreau*}

\affil[1]{Quantum and Nanotechnologies Research Centre, National Research Council Canada, Ottawa, Ontario, K1A 0R6, Canada}
\affil[2]{Department of Physics, University of Ottawa, Ottawa, Ontario, K1N 9A7, Canada}
\affil[3]{Research Center for Materials Nanoarchitectonics, National Institute for Materials Science, 
Tsukuba, Japan}
\affil[4]{Research Center for Electronic and Optical Materials, 
National Institute for Materials Science, Tsukuba, Japan}
\affil[5]{Department of Inorganic Chemistry, University of Chemistry and Technology Prague, Technicka 5, 166 28 Prague 6,Czech Republic}
\affil[6]{University of Illinois Chicago, Chicago, USA}
\title{Establishing  the role of thickness in LaOBr; an alternative gate dielectric for vdW heterostructures}
\date{*Louis.Gaudreau@nrc-cnrc.gc.ca}

\begin{document}

\maketitle

\begin{abstract}
In electronic devices, the quality of the dielectric material is determined by figures of merit such as the band gap, leakage onset field, breakdown field and the number of built-in defects, which can all limit device performance. An additional and important metric to consider is the strength of the dielectric constant. Hexagonal boron nitride (hBN) is the benchmark gate dielectric for two-dimensional van der Waals heterostructures, owing to its wide bandgap, and atomically flat interface. At the same time, its dielectric constant (2--4) highlights the opportunity to explore alternative higher-permittivity van der Waals dielectrics that provide stronger electrostatic control. Previous work established lanthanum oxybromide (LaOBr), a layered dielectric with a predicted dielectric constant of 12.5, as a potential dielectric for van der Waals heterostructures. Using graphene field-effect devices with LaOBr, we show that the leakage-onset and breakdown fields decrease systematically with thickness. Dual-gate capacitance measurements combined with a dead-layer model reveal a bulk dielectric constant of $\varepsilon_{\mathrm{bulk}} = 12.2$, in agreement with theory, while interfacial dead-layers cause the effective dielectric constant to vary from 6 to 11 across multiple thicknesses. Benchmarked against established gate dielectrics, LaOBr combines breakdown fields comparable to hBN with dielectric constants 2 to 3 times larger. By establishing the operating regime of devices with various thicknesses, we further confirm the advantage of LaOBr over hBN for high-capacitance gate-dielectric operation. These results validate LaOBr as a promising dielectric for integration into two-dimensional heterostructures, and establish thickness-dependent benchmarking as essential for evaluating emerging layered dielectric materials.
\end{abstract}

\section*{Keywords}

two-dimensional materials, dielectrics, van der Waals heterostructures

\section{Introduction}
Two-dimensional (2D) van der Waals (vdW) heterostructures have enabled the study of new atomically thin materials for electronic, spintronic and optoelectronic applications, allowing the exploration of a remarkable variety of correlated, topological and moiré-driven phases of matter \cite{geim2013van,ren20262d,ajayan2016two}. The performance of these devices is linked to the quality of its gate dielectric. The dielectric does far more than merely insulate: it determines the maximum carrier density that can be electrostatically induced, screens the electric field, imposes the capacitance, and can influence device properties, such as its mobility, which are dependent on the quality of the dielectric's interface with the material \cite{dean2010boron,}. An ideal 2D dielectric should therefore provide among other desirable characteristics, the following four properties : a wide band gap to suppress leakage, a high dielectric constant to maximize the gate capacitance, a high breakdown field to allow strong electrostatic doping without catastrophic failure, and atomically clean van der Waals  interfaces for compatibility with 2D heterostructure assembly \cite{yang2023gate}.
\\
\\
Among all the parameters that govern these figures of merit, the thickness of the dielectric layer is a very important design knob. For example, in field-effect devices, it controls the gate capacitance,  and sets the operating voltage. Additionally, it dictates the dominant leakage mechanism. At the thinnest scales, leakage is governed by direct tunneling through the dielectric. For thicker films, field effect emission and trap-assisted transport mechanisms such as the Poole–Frenkel effect determines the leakage current. \cite{yuan2026quantum}. However, these properties need not follow simple or universal trends with thickness. Consequently, establishing the performance of a dielectric material requires more than a single-thickness measurement. A systematic thickness-dependent study is necessary to determine how the dielectric constant, leakage behaviour, and breakdown strength evolve with layer number, and to identify the regime in which the material offers advantages over conventional dielectrics.
\\
\\
Hexagonal boron nitride (hBN) has so far been the benchmark 2D dielectric in vdW heterostructures \cite{illarionov2020insulators,https://doi.org/10.1002/adma.202207901,,10.1039/c8cs00450a}. It satisfies the cleanliness and bandgap criteria but its modest dielectric constant (reported from 2 to 4 \cite{laturia2018dielectric,kim2012synthesis,young2012electronic,https://doi.org/10.1002/adfm.201804235}) limits the gate capacitance. Exploring alternative layered dielectrics and studying their performance as a function of thickness is therefore the natural route to advance 2D devices beyond hBN \cite{Li2026,doi:10.1021/acsnano.3c10411,Li2025,Holler2020,Wang2022,Li2021,Jia2026,fu2025low}
\\
\\
In this work, we focus on Lanthanum Oxybromide (LaOBr), a 2D dielectric material with a wide band gap of about 5.3 eV \cite{doi:10.1021/acsnano.3c10411} and an out-of-plane dielectric constant of 12.5, as predicted by density functional theory \cite{osanloo2021identification}. We investigate how its key electrical figures of merit, such as the leakage current, dielectric constant, leakage-onset field $E_{\mathrm{LC}}$, and breakdown field $E_{\mathrm{BD}}$, scale with thickness. The central finding is that the thickness dependence itself uncovers a regime in which LaOBr combines a dielectric constant  2 to 3 times larger than hBN with a comparable breakdown field.  Beyond this specific material, our results show that a thickness-dependant comparison is indispensable to fairly evaluate any new 2D dielectric and to identify the range in which it can genuinely advance the performance of van der Waals devices.

\section{Results and discussion}

\subsection{Structural and spectroscopic characterization of LaOBr}
\begin{figure*}[t]
  \centering
    \includegraphics[width=\linewidth]{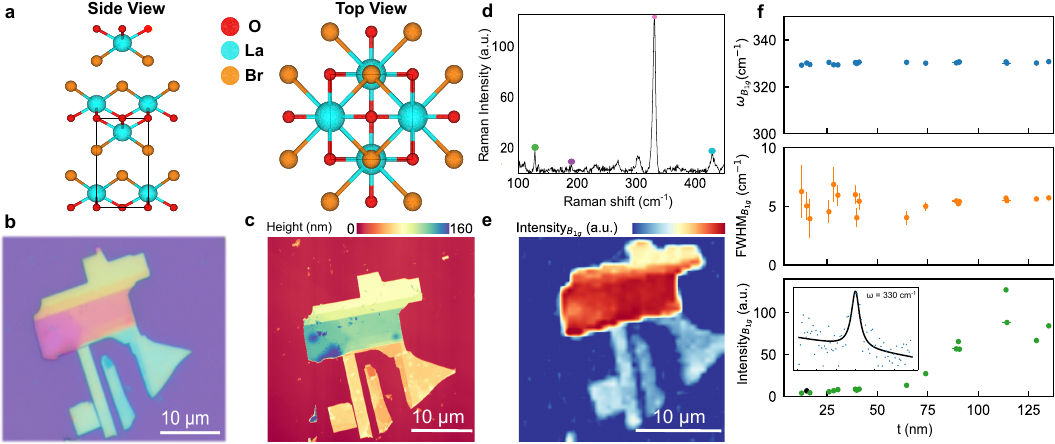}
    \caption{\textbf{Structural and spectroscopic characterization of LaOBr.} \textbf{(a)} Crystal structure of LaOBr visualized with VESTA \cite{https://doi.org/10.1107/S0021889811038970}, showing the tetragonal symmetry with La in cyan, O in red, and Br in orange. \textbf{(b)} Optical micrograph  of a LaOBr flake showing thickness-dependent color contrast. \textbf{(c)} Atomic force microscopy (AFM) image confirming the thickness variations observed in \textbf{(b)}. \textbf{(d)}  Raman spectrum of a LaOBr flake showing the different peaks at 128 ($A^1_{1g}$), 189 ($A^2_{1g}$), 330 ($B_{1g}$), and 428  cm$^{-1}$ ($E^2_{g}$) respectively represented with a green, purple, pink and blue dot.   \textbf{(e)} Raman intensity map of the flake at the  $B_{1g}$ peak position, revealing spatial variations in Raman intensity. \textbf{(f)} Thickness dependence of the $B_{1g}$ mode: peak frequency, FWHM, and intensity. The inset shows the Raman intensity as a function of Raman shift for the $B_{1g}$ peak of a 15-nm-thick flake (black dot), confirming the presence of the mode for thinner flakes even at the lowest intensities. The black curve is the  Lorentzian fit of the experimental data points shown in blue.}
    \label{fig1}
\end{figure*}

We start our study by establishing the crystal structure and exfoliation quality of LaOBr. Figure \ref{fig1}a shows the crystal structure visualized with VESTA \cite{https://doi.org/10.1107/S0021889811038970}; lanthanum atoms are rendered in cyan, oxygen in red, and bromine in orange. The side and top views  reveal a tetragonal crystal structure, distinguishing it from commonly used van der Waals materias such as hBN and WSe$_2$, which possess hexagonal lattices. This difference in lattice symmetry may provide an appealing route for LaOBr-based heterostructures, as symmetry-mismatched van der Waals interfaces have been shown to generate moiré potentials with reduced symmetry compared with the constituent layers \cite{Sajan2026}.
\\
\\
Figure \ref{fig1}b presents an optical micrograph of an exfoliated LaOBr flake which displays thickness-dependent color contrast. Thin locations appear blue whereas thicker regions show pink tones. Figure \ref{fig1}c shows the corresponding atomic force microscopy image, which confirms a layered-step structure  with well defined boundaries between regions of different thickness. The measured thickness spans from a few nanometers up to 160 nm, demonstrating that multi-thickness LaOBr flakes can be obtained by mechanical exfoliation.
\\
\\
To further characterize the flake, we performed  Raman spectroscopy  and identified four different peaks originating from the LaOBr flake (Figure \ref{fig1}d), the most intense one at around 330 cm$^{-1}$ associated to the B$_{1g}$ mode, an antisymetric out-of-plane mode \cite{doi:10.1021/acsnano.3c10411}. To assess the evolution of its vibrational response with thickness, we performed spatially resolved Raman spectroscopy. Figure \ref{fig1}e displays a Raman intensity map recorded at the B$_{1g}$ peak associated to the flake in Figure 1b. The map reveals the  thickness dependence of this mode: its intensity is weak on the thinnest areas and increases with thickness, mirroring the topology obtained with the AFM. This behavior is quantified in Figure \ref{fig1}f  where we plot the peak position, full-width at half-maximum (FWHM) and the  intensity of the B$_{1g}$ peak as a function of thickness. The peak position and FWHM remain essentially independent of  thickness, confirming that the  B$_{1g}$ vibrational mode and hense the LaOBr tetragonal crystal structure is preserved across the entire range investigated.

\subsection{Electrical characterization of LaOBr as a gate dielectric}
\begin{figure*}[t]
  \centering
    \includegraphics[width=\linewidth]{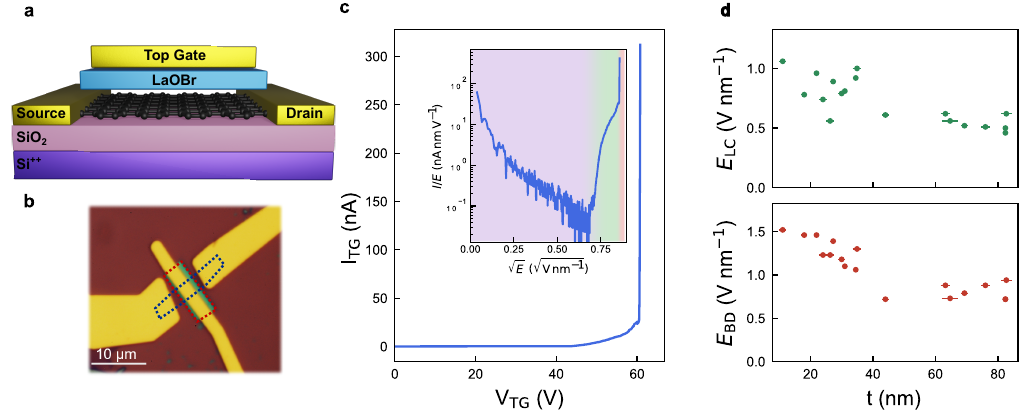}
    \caption{\textbf{Electrical characterization of LaOBr as a gate dielectric}. \textbf{(a)} Schematic of the graphene field-effect device used for dielectric characterization. From bottom to top: p-doped Si serves as the back-gate; a SiO$_2$  layer supports a graphene flake contacted by source and drain electrodes; a LaOBr flake is transferred on the channel and is capped by a metallic top-gate. \textbf{(b)} Optical micrograph of a characterized device. LaOBr highlighted in red and graphene in blue. \textbf{(c)} Top-gate leakage current I$_{\mathrm{TG}}$ as a function of  $\mathrm{V_{TG}}$ . Inset:
    Frenkel-Poole plot $\log(\mathrm{I_{TG}}/E)$  versus $\sqrt{E}$,  used to delimit the operating regime (purple), from the leakage-onset regime (green) preceding the breakdown regime (red). \textbf{(d)} Thickness dependence of the leakage-onset field $E_{\mathrm{LC}}$ (top) and breakdown field $E_{\mathrm{BD}}$ (bottom) for LaOBr flakes with thicknesses ranging from 11.1 nm to 82.7 nm.}
    \label{fig2}
\end{figure*}
To assess the suitability of LaOBr as a gate dielectric for van der Waals heterostructure devices, we fabricated graphene field-effect devices in which an exfoliated layer of LaOBr serves as the top-gate dielectric (Figure \ref{fig2}a). The stack consists of graphene mechanically exfoliated onto a 285 nm SiO$_2$ / Si$^{++}$ substrate, and contacted by Ti/Pd/Au source and drain electrodes. A LaOBr flake of thickness t is transferred on top of the graphene using a PPC/PDMS stamp, followed by the deposition of a metallic top-gate. An optical micrograph of a device is shown in (Figure \ref{fig2}b). In total, 18 devices were fabricated and measured, with LaOBr thicknesses ranging from 11 nm to 83 nm.
\\
\\
We first characterized the vertical transport across the LaOBr layer by varying the  voltage $\mathrm{V_{TG}}$ while monitoring the gate leakage current I$_{\mathrm{TG}}$ (Figure \ref{fig2}c). The leakage current remains below the floor level (<10 pA) before rising and ultimately diverging at the dielectric breakdown. To identify  the underlying leakage mechanisms, we reploted the data over the entire voltage range as a Poole–Frenkel diagram, $\log(\mathrm{I_{TG}}/E)$ as a function of $\sqrt{E}$ , shown as an inset of the main plot of Figure \ref{fig2}c, where $E = \mathrm{V_{TG}}/\mathrm{t}$ is  the vertical electric field field across the LaOBr layer.
\\
\\
Three distinct regimes emerge and are highlighted by color in Figure \ref{fig2}c. Optimal device performance requires working within the operating regime (purple) where no parasitic effects come from leakage currents, a leakage onset regime (green) in which log(I$_{\mathrm{TG}}/E)$ increases with  $\sqrt{E}$,
and a breakdown regime (red) in which the current diverges.
\\
\\
To benchmark LaOBr as a dielectric, we extracted two metrics from each device: the leakage-onset field $E_{\mathrm{LC}}$, i.e. the field at which I$_{\mathrm{TG}}$ starts increasing, and the breakdown field $E_{\mathrm{BD}}$ defined as the field at which the current diverges. Figure \ref{fig2}d illustrates how $E_{\mathrm{LC}}$ (top) and $E_{\mathrm{BD}}$ (bottom) depend on the LaOBr flake thickness for 18 different devices. Both $E_{\mathrm{LC}}$ and $E_{\mathrm{BD}}$ decrease as t increases, varying from $1.06$ to $0.46~\mathrm{V\,nm^{-1}}$ and from $1.52$ to $0.72~\mathrm{V\,nm^{-1}}$. This thickness dependence is consistent with a defect-mediated breakdown scenario in which the probability of finding a  percolation path through the dielectric grows with the insulator volume, a behavior reported for  traditional high-$\kappa$
dielectrics \cite{lombardo2005dielectric}  and also, more recently, hBN \cite{sheng2025dielectric,10.1063/1.1611649}.
\subsection{Dielectric constant extraction and dead-layer model.}
\begin{figure*}[t]
  \centering
    \includegraphics[width=\linewidth]{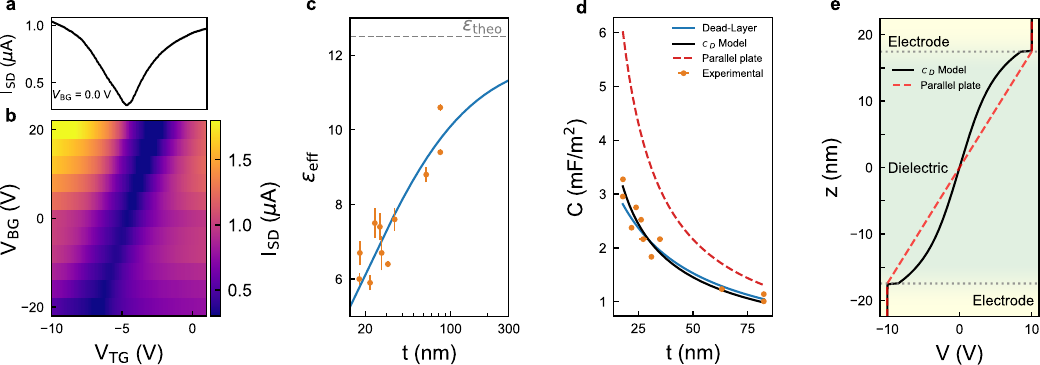}
    \caption{\textbf{Dielectric constant extraction and dead-layer model.} \textbf{(a)} Source-drain current $\mathrm{I_{SD}}$ as a function of top-gate voltage $\mathrm{V_{TG}}$ taken at V$_{\mathrm{BG}} = 0$ V and source-drain bias V$_{\mathrm{SD}} = 5$ mV. The Dirac point corresponds to the top-gate voltage where the current is at its minimum. \textbf{(b)}  $\mathrm{I_{SD}}$ as a function of  $\mathrm{V_{TG}}$ and $\mathrm{V_{BG}}$ taken at V$_{\mathrm{SD}}=5$ mV. The shift of the Dirac point with respect to $\mathrm{V_{BG}}$ provides the relative gate capacitance between the top and back gates. \textbf{(c)} Effective dielectric constant $\varepsilon_{\mathrm{eff}}$ extracted from the measured capacitance as a function of LaOBr thickness t (orange points). The dashed grey line marks the theoretical bulk value $\varepsilon_{\mathrm{theo}}$  = 12.5. The blue curve is a fit to the experimental data using the dead-layer model, which converges to $\varepsilon_{\mathrm{bulk}} = 12.2$ in the limit t $\xrightarrow{}\infty$. \textbf{(d)} Capacitance per unit area of the LaOBr dielectric as a function of flake t. Orange circles: experimental data  from panel c. Red dashed curve: parallel plate capacitor model, $C_\mathrm{{pp}} = \varepsilon_{\mathrm{bulk}}\varepsilon_0/\mathrm{t}$, with $\varepsilon_{\mathrm{bulk}}= 12.2 $. Solid blue curve: capacitance reconstructed from the dead-layer model (Eq.~\ref{dead2}) . Solid black curve: fit to the flexoelectric dead-layer capacitance expression from \cite{majdoub2009understanding}, yielding a longitudinal flexoelectric length scale l = 4.5 nm and a field-penetration parameter k = 0.35. \textbf{(e)} Electric potential distribution across a capacitor. The red dashed curve corresponds to the parallel plate capacitor model, whereas the black curve accounts for the dead-layer model.}
    \label{fig3}
\end{figure*} 
To quantify the dielectric response of LaOBr, we performed dual-gate transport measurements by applying a displacement field between the top-gate and the Si$^{++}$ back-gate. Figure \ref{fig3}a shows a representative line-cut of the  current through the graphene channel  $\mathrm{I_{SD}}$ as a function of  $\mathrm{V_{TG}}$ taken at a fixed V$_{\mathrm{BG}} = 0$ V, where the dip in current corresponds to the Dirac point of graphene, reached under the combined effect of the top and back-gates. Figure \ref{fig3}b is a color map of $\mathrm{I_{SD}}$ as a function of $\mathrm{V_{TG}}$ and $\mathrm{V_{BG}}$, obtained by repeating this measurement for successive values of $\mathrm{V_{BG}}$. The charge neutrality point of the graphene channel follows a diagonal line in the $\mathrm{V_{TG}}$--$\mathrm{V_{BG}}$ plane, where its slope is set by the ratio of the two gate capacitances: 
\begin{equation}
\frac{C_{\mathrm{SiO_2}}}{C_{\mathrm{LaOBr}}}
=
\frac{\mathrm{\varepsilon}_{\mathrm{SiO_2}}\, \mathrm{t_{\mathrm{LaOBr}}}}
{\mathrm{\varepsilon}_{\mathrm{LaOBr}}\, \mathrm{t_{\mathrm{SiO_2}}}}
\label{cap}
\end{equation}
\\
Using the SiO$_2$ thickness and dielectric constant  (t$_{\mathrm{SiO_2}} = 285$ nm, $\varepsilon_{\mathrm{SiO_2}} = 3.9$ \cite{Robertson_2004})  and the AFM-measured LaOBr thickness, this expression allows us to extract  the effective out-of-place dielectric constant ($\varepsilon_{\mathrm{eff}}$) of each LaOBr flake. The resulting $\mathrm{\varepsilon_{eff}}$ is plotted as a function of t in Figure \ref{fig3}c. Rather than being a thickness-independent material constant, $\varepsilon_{\mathrm{eff}}$ is found to increase with t, rising from  5.9 to 10.6,  and approaches the DFT-predicted dielectric constant of 12.5 (grey dashed line). This behavior is a direct  consequence of the dead-layer effect \cite{stengel2006origin,majdoub2009understanding}. In this framework, each electrode--dielectric interface is assumed to host a thin region of reduced permittivity, known as the dead-layer. The effective capacitance of the resulting structure can then be represented by three capacitive elements in series: the two dead-layers and the ideal bulk dielectric capacitance. The total capacitance is then given by:
\begin{equation}
   \frac{1}{\mathrm{C}_{\mathrm{eff}}} = \frac{1}{\mathrm{C}_{\mathrm{bulk}}} + \frac{1}{\mathrm{C}_{i_1}} + \frac{1}{\mathrm{C}_{i_2}}
\label{dead1}
\end{equation}
Here, $\mathrm{C_{eff}}$ is the total effective capacitance, $\mathrm{C_{bulk}}$ is the bulk dielectric capacitance, and $\mathrm{C}_{i_1}$ and $\mathrm{C}_{i_2}$ are the capacitances associated to the two dead-layers. Assuming that the interfacial dead-layer thicknesses t$_{i_1}$ and t$_{i_2}$ are independent of the total flake thickness, Equation~\ref{dead1} can be rewritten as a linear relation between t$/\varepsilon_{\mathrm{eff}}$ and t \cite{huang2022high}:
\\
\begin{equation}
  \frac{\mathrm{t}}{\varepsilon_{\mathrm{eff}}} = \frac{\mathrm{t}}{\varepsilon_{\mathrm{bulk}}} + D, \qquad D \equiv \mathrm{t_{i_1}}\left(\tfrac{1}{\varepsilon_{i_1}}-\tfrac{1}{\varepsilon_{\mathrm{bulk}}}\right) + \mathrm{t_{i_2}}\left(\tfrac{1}{\varepsilon_{i_2}}-\tfrac{1}{\varepsilon_{\mathrm{bulk}}}\right)
\label{dead2}
\end{equation}
\\
The thickness-independent bulk dielectric constant $\varepsilon_{\mathrm{bulk}}$ is obtained by fitting our data to equation \ref{dead2} represented by the blue curve in Figure \ref{fig3}c, where $D$ directly quantifies the total dead-layer contribution, resulting in $\mathrm{\varepsilon_{bulk} = 12.2}$.
\\
\\
In a similar manner, the experimentally measured capacitances are shown as orange dots  in Figure \ref{fig3}d as a function of dielectric thickness. Using the fitted bulk dielectric constant, $\varepsilon_{\mathrm{bulk}} = 12.2$, together with the total dead-layer contribution, $D$, extracted from Figure \ref{fig3}c, we calculate the expected capacitance including interfacial dead-layer effects, shown as the solid blue curve. This thickness dependence can be directly compared with the  parallel plate capacitor model, shown as the red dashed curve, which assumes a uniform dielectric constant and neglects the interfacial dead-layers. Finally, following the theoretical approach in Ref. \cite{majdoub2009understanding} where they consider flexoelectricity, a strain gradient-induced polarization, to improve the understanding of the dead-layer effect, we further analyze the thickness-dependent capacitance using the expression:

\begin{equation}
    \mathrm{C}_D = \mathrm{C}_{\mathrm{pp}} \frac{1+(\mathrm{k}(\varepsilon_{\mathrm{bulk}}-1)\mathrm{l}/\mathrm{t})\mathrm{tanh}(\mathrm{t}/\mathrm{l})}{1+((\varepsilon_{\mathrm{bulk}}-1)\mathrm{l}/\mathrm{t})\mathrm{tanh}(\mathrm{t}/\mathrm{l})}
\label{Cd}    
\end{equation}
\\
\\
Here, C$_D$ is the capacitance of the dielectric material including interfacial dead-layer effects, C$_{\mathrm{pp}}$ is the parallel plate capacitance, k characterizes the electric-field penetration, and l is the flexoelectric length scale. Fitting our experimental data to this expression gives the black curve in Figure \ref{fig3}d, yielding k $ = 0.35$ and l $ = 4.5$ nm.
\\
\\
To better visualize the influence of the interfacial dead-layer, Figure \ref{fig3}e plots the potential distribution across the capacitor. The red dashed curve corresponds to the parallel plate capacitor picture, while the black curve shows the modified potential profile obtained when dead-layer effects are included.
\\
\\
Two conclusions follow: first, the intrinsic dielectric constant of LaOBr, $\varepsilon_{\mathrm{bulk}} = 12.2$, is markedly larger than that of hBN ($\varepsilon_{\mathrm{hBN}}$ = 2-4) and places LaOBr in the family of promising high-$\kappa$ van der Waals dielectrics. Second, despite accounting for the dead-layer penalty, the effective dielectric constant of LaOBr, measured over a wide range of thicknesses, remain consistently larger than $\varepsilon_{\mathrm{hBN}}$.

\subsection{Benchmarking LaOBr against established gate dielectrics.}
\begin{figure*}[t]
  \centering
    \includegraphics[width=\linewidth]{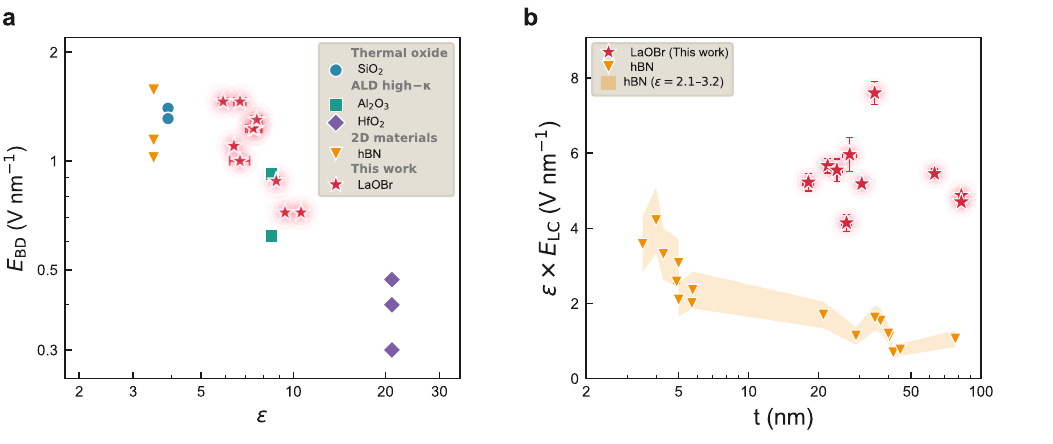}
    \caption{\textbf{Benchmarking LaOBr against established gate dielectrics.} \textbf{(a)} Breakdown field $E_{\mathrm{BD}}$ as a function of dielectric constant $\varepsilon$ of representative gate insulators: thermal SiO$_2$ (blue circles), ALD high-$\kappa$ oxides Al$_2$O$_3$ (green squares) and HfO$_2$ (purple diamonds), 2D hBN (orange triangles), and LaOBr (red stars). LaOBr data points are a result of this study while other data points are taken from \cite{McPherson2002ProposedUR,wang2025dielectric}. \textbf{(b)} $\varepsilon \times E_{\mathrm{LC}}$ as a function of dielectric thickness  for LaOBr (red stars) and hBN (orange triangles). }
    \label{fig4}
\end{figure*}

To place LaOBr in the broader landscape of gate dielectrics, Figure \ref{fig4}a compares the breakdown field $E_{\mathrm{BD}}$ against the dielectric constant $\varepsilon$ for the most widely used insulators in modern field-effect devices: the thermal oxide SiO$_2$, the  high-$\kappa$ oxides grown by Atomic Layer Deposition (ALD), Al$_2$O$_3$ and HfO$_2$, and the 2D reference dielectric hBN. Each red star corresponds to individual LaOBr devices with different thicknesses measured in this work, while the literature values are taken from \cite{McPherson2002ProposedUR, wang2025dielectric}.
\\
\\
From Figure \ref{fig4}a a trade-off emerges across all material families: dielectrics with larger $\varepsilon$ tend to exhibit lower $E_{\mathrm{BD}}$, a trend rooted in the inverse correlation between bandgap and polarizability \cite{McPherson2002ProposedUR}.
SiO$_2$ and hBN sit at the low-$\varepsilon$/high $E_{\mathrm{BD}}$, while HfO$_2$ occupies the opposite corner  ($\varepsilon$ $\approx$ 21, $E_{\mathrm{BD}}\approx 0.3-0.5$ V nm$^{-1}$) and Al$_2$O$_3$  falls in between ($\varepsilon$ $\approx$ 0.5, $E_{\mathrm{BD}}\approx 0.6-0.9$ V nm$^{-1}$).
\\
\\
LaOBr provides a favorable balance between dielectric response and breakdown strength. Across the thickness range studied here, LaOBr exhibits dielectric constants between approximately 6 and 11, while maintaining breakdown fields up to $\approx 1.5$ V nm$^{-1}$. This combination places LaOBr at a higher dielectric constant than low-$\varepsilon$ dielectrics such as SiO$_2$ and hBN, while retaining breakdown fields of comparable magnitude. In this respect, LaOBr provides an improved dielectric response at similar breakdown fields, highlighting its potential as a robust van der Waals dielectric for electrostatically gated devices.
\\
\\
While the breakdown field $E_{\mathrm{BD}}$  shown in  Figure \ref{fig4}a  sets the ultimate failure limit of the dielectric, optimal gate operation must remain below this value to avoid leakage currents. A more relevant figure of merit is therefore the maximum electric field sustainable within the operating regime, defined here by the leakage current onset $E_{\mathrm{LC}}$, weighted by the dielectric constant, $\varepsilon \times E_{\mathrm{LC}}$. This figure of merit is plotted in Figure \ref{fig4}b as a function of thickness for our LaOBr devices. For comparison, we fabricated devices using hBN as the gate dielectric and characterized following the same procedure; these data are shown as orange triangles. This quantity reflects the maximum gate-induced charge density achievable before leakage becomes significant, and thus provides a practical measure of gate efficiency under operating conditions. As shown in Figure \ref{fig4}b, LaOBr exceeds hBN across the explored thickness range. 

\subsection{CONCLUSION}
Taken together with the low leakage current (Figure \ref{fig2}c) and the high $\varepsilon \times E_{\mathrm{LC}}$ within the operating regime (Figure \ref{fig4}), these results establish LaOBr as a promising  2D gate dielectric, offering a combination of high capacitance, high breakdown strength, and full compatibility with van der Waals heterostructures. This positions LaOBr as an  alternative dielectric material in  next-generation 2D electronics where both gate efficiency and dielectric robustness are required.
LaOBr's electronic properties make it a interesting choice as a building block for 2D-material based devices, where  the natural thinness of van der Waals crystals, combined with  the wide tunability of the dielectric constant we report here, opens a direct knob to engineer electrostatic screening and modulate interactions. Adding this work  with  recent publication   \cite{doi:10.1021/acsnano.3c10411,Li2026} we  demonstrate that LaOBr can be successfully integrated and operated within functional van der Waals devices, providing concrete proof that the material is compatible with the standard 2D-heterostructure fabrication.

\subsection{METHODS}
\subsubsection{Structural and spectroscopic characterization of LaOBr}

Raman spectroscopy was performed with a commercial microRaman system (LabRAM HR, Horiba Jobin Yvon, France) equipped with a computer controlled XYZ stage and a thermoelectric cooled CCD detector and a 80-cm focal length spectrometer. The instrument is calibrated to 520.7 cm$^{-1}$ peak of a Si wafer. Samples were excited with 457.12 nm radiation at a power of 1.6 mW. Incident radiation was coupled into an Olympus BX51 optical microscope and focused down to an $\approx$ 1~\textmu m diameter spot through a 100X objective (0.9 NA). Raman spectra were acquired with 1800 g/mm grating and 2 s / 10 scan acquisition time. For Raman mapping measurements, the samples are raster under the focused laser beam and scanned in a grid pattern with a constant 0.7 µm steps in both x and y directions.

\subsubsection{Device fabrication}
All devices were  fabricated on  a Si$^{++}$/SiO$_2$ substrates. LaOBr flakes were exfoliated onto a substrate and transferred on top of the graphene using a Polypropylene Carbonate (PPC) / Polydimethylsiloxane (PDMS) stamp. The contacts and top-gates were fabricated following standard electron-beam lithography procedures, followed by metal deposition of Ti (50 \AA) + Pd (200 \AA) + Au (1200 \AA).

\subsubsection{Electrical measurements}
Electrical measurements were performed at 1.5 K in a dry ICE Oxford cryostat using Keysight B2911 source measure units. A HP 3245A universal source was used to apply a voltage to the top-gate.

\section*{Acknowledgements}
 A.L.-M and L.G. acknowledge support from the the Quantum Sensors Challenge Program at the National Research Council of Canada. A.L.-M., and L.G. acknowledge support from the Natural Sciences and Engineering Research Council of Canada (NSERC) AllRP/578466-2022, and NSERC CREATE/ 575280-2023 - Training in Materials for Quantum Technologies (MaQTech). K.W. and T.T. acknowledge support from the CREST (JPMJCR24A5), JST and World Premier International Research Center Initiative (WPI), MEXT, Japan. Z.S. were supported by project LUAUS25268 funded by Ministry of Education, Youth and Sports of the Czech Republic.

\printbibliography


\newpage

\end{document}